\documentclass[11pt]{article}
\usepackage[margin=1in]{geometry}
\usepackage{graphicx}
\usepackage{booktabs}
\usepackage{subcaption}
\usepackage{amsmath}
\usepackage{needspace}
\usepackage{placeins}
\usepackage{xurl} 
\usepackage[hidelinks]{hyperref}

\title{WritePolicyBench: Benchmarking Memory Write Policies under Byte Budgets}
\author{Edgard El Cham}
\date{}

\begin{document}
\maketitle

\begin{abstract}
We introduce WritePolicyBench, a benchmark for evaluating \emph{memory write policies}---decision rules that choose what to store, merge, and evict under a strict byte budget while processing a stream with document/API drift. The benchmark provides (i) task generators with controlled non-stationarity, (ii) an explicit action interface for external memory, (iii) a byte-accurate cost model, and (iv) standardized metrics that measure both task success and budget efficiency.
\end{abstract}

\section{Introduction}
Bounded memory is a core constraint in streaming systems, long-context agents, and tools that maintain external state. When the input stream is larger than the available byte budget, overall performance depends not only on retrieval but also on \emph{what gets written}. A \emph{write policy} must decide which observations to store, when to update existing items, and what to evict so that the memory retains information that remains useful under distribution drift.

\paragraph{Research question.} \emph{How do different memory write policies trade task success, budget utilization, and robustness to non-stationarity when constrained by a strict byte budget?}

WritePolicyBench is a benchmark designed to isolate and measure write-policy behavior under controlled synthetic drift. The benchmark provides:
\begin{itemize}
  \item task generators with parameterized drift regimes (default, burst drift, and redundancy);
  \item an explicit action interface for external memory (WRITE/EXPIRE and optional delta updates);
  \item a byte-accurate cost model with standardized evaluation metrics that capture both task quality (e.g., F1, regret) and memory behavior; and
  \item a reproducible evaluation pipeline with deterministic episodes and reference baselines.
\end{itemize}

The remainder of the paper specifies the benchmark (Sections~\ref{sec:benchmark}--\ref{sec:metrics}) and reports baseline results that illustrate budget tradeoffs and characteristic failure modes (Sections~\ref{sec:experiments}--\ref{sec:results}).

\section{Benchmark}
\label{sec:benchmark}
WritePolicyBench evaluates policies that process a stream of steps $t=1,\dots,T$. Each step provides an observation $x_t$ and optional metadata $m_t$. The policy maintains an external memory $M_t$ whose contents must fit within a fixed byte budget $B$.

The benchmark isolates the \emph{write problem}: deciding \emph{what to store} (and when to merge/evict) under a strict budget so that relevant drift information is retained.

\paragraph{Episode schema.} Each episode consists of:
(i) a sequence of steps $\{(x_t, m_t)\}_{t=1}^T$;
(ii) hidden labels that define a set of relevant/critical timesteps $R$ (e.g., breaking-change events);
and (iii) a hidden per-step utility $u_t$ that rewards retaining high-value events.

\paragraph{Policy observability and evaluation tracks.}
We distinguish what is used for \emph{scoring} from what is \emph{observable} to the policy.
In the \textbf{unprivileged} track, the policy observes only $(x_t, m_t)$, where $m_t$ is restricted to benign metadata (e.g., episode regime identifiers) and excludes labels and utility.
In the \textbf{privileged} track, the policy additionally observes a scalar \texttt{priority} signal $p_t$ (a bounded surrogate correlated with $u_t$).
Both tracks are scored against the same hidden labels $R$ and utility $u_t$.

\paragraph{Policy interface.} At each step, the policy emits a sequence of memory actions (Section~\ref{sec:memiface}). Actions are applied with budget enforcement and semantic constraints (e.g., EXPIRE targets must be older than the current timestep).

\paragraph{What is scored.} After processing the full stream, we score the final retained set of timesteps $W$ against the labels $R$ using precision/recall/F1 and additional budget-efficiency metrics (Section~\ref{sec:metrics}).

\section{Task generation: synthetic drift regimes}
\label{sec:taskgen}
We provide synthetic generators designed to make relevance observable and to create controlled forms of non-stationarity. Each generator emits a stream of API ``snapshots'' where endpoint versions and parameter sets may drift over time. Certain steps are labeled as \emph{critical} (e.g., breaking-change events), and a per-step utility $u_t$ rewards retaining information about such events.

We include multiple regimes:
\begin{itemize}
  \item \textbf{Default:} drift events occur i.i.d. at a base rate.
  \item \textbf{Burst drift:} drift probability increases sharply during short windows, producing clusters of breaking changes.
  \item \textbf{Redundancy:} the stream contains repeated or near-duplicate observations, so naive write-all policies waste budget on duplicates.
  \item \textbf{Burst+redundancy:} combines clustered drift with redundant streaks, stressing both burst robustness and duplicate filtering.
\end{itemize}

The key design choice is that relevance is defined from the generator itself (via labels such as \texttt{critical\_steps} and \texttt{total\_drift\_events}), enabling deterministic evaluation without an LLM-in-the-loop.

\section{Memory interface}
\label{sec:memiface}
We model memory as an append-only store with explicit actions and a fixed byte budget $B$.
At each step $t$, a policy observes the current step $(x_t, m_t)$ and the current memory state $M_t$ (including byte usage), then emits a sequence of actions from:
\begin{itemize}
  \item \textbf{WRITE(step):} store the full step, charging \texttt{estimate\_bytes(step)}.
  \item \textbf{MERGE(target, delta):} append a delta update into an existing \emph{base} item (delta-only), charging $\texttt{bytes(delta)} + 16$.
  \item \textbf{EXPIRE(target):} remove a stored item and immediately credit its original byte cost.
  \item \textbf{SKIP:} no-op.
\end{itemize}

\paragraph{Constraints.} Actions that would exceed budget are rejected. EXPIRE is only valid for items older than the current timestep (preventing trivial write-then-expire budget circumvention).

MERGE is constrained to prevent degenerate compression strategies: it may only target a \emph{base WRITE} item (no MERGE-to-MERGE chains), may only apply within the same endpoint identity (matching the observation's \texttt{api} field), and the delta must be the canonical shallow field-diff excluding \texttt{api} and must be non-empty (no-op merges are rejected). The implementation follows the benchmark specification's byte model, including per-item index overhead.

\section{Baselines}
\label{sec:baselines}
We include simple, deterministic baselines to (i) sanity-check the evaluation pipeline and (ii) provide reference curves for budget tradeoffs.

\paragraph{Unprivileged track.} These policies use only the observable stream $(x_t, m_t)$ (no labels/utility/priority).
\begin{itemize}
  \item \textbf{No memory (\texttt{no\_mem}):} always SKIP.
  \item \textbf{Fill-then-stop (\texttt{fifo\_store\_all}):} WRITE whenever the remaining budget can accommodate the current step; otherwise SKIP (no eviction).
  \item \textbf{Last-$k$ bytes (\texttt{last\_kb}):} before writing, EXPIRE the oldest items until the current step fits.
  \item \textbf{Uniform sample (\texttt{uniform\_sample}):} deterministically write every $N$-th step (we use $N{=}10$).
  \item \textbf{Merge-aggressive (\texttt{merge\_aggressive}):} when an item for the same API exists, MERGE deltas into the most recent match; otherwise fall back to \texttt{last\_kb}.
\end{itemize}

\paragraph{Privileged track.} These policies additionally observe a scalar \texttt{priority} hint $p_t$.
\begin{itemize}
  \item \textbf{Priority threshold (\texttt{priority\_threshold}):} WRITE only steps whose \texttt{priority} metadata exceeds a threshold (we use $\tau{=}0.5$).
  \item \textbf{Priority-greedy (\texttt{priority\_greedy}):} maintain a set of high-priority items by evicting the lowest-priority items to make room (ties broken by evicting the oldest item).
\end{itemize}

\section{Metrics}
\label{sec:metrics}
We report task-quality metrics and memory-efficiency metrics computed from per-episode logs.

\paragraph{Task quality.} Each episode defines a set of relevant timesteps $R$ (e.g., breaking changes). A policy yields a set of retained timesteps $W$ induced by the final memory contents. WRITE items contribute their timestep to $W$. MERGE delta items contribute their timestep only if they reference a still-present base WRITE item of the same endpoint identity (matching \texttt{api}); orphan deltas (whose base was expired) are not counted. We compute:
\begin{align}
\text{precision} &= \frac{|W \cap R|}{|W| + \epsilon}, \\
\text{recall} &= \frac{|W \cap R|}{|R| + \epsilon}, \\
\text{F1} &= \frac{2\,\text{precision}\cdot\text{recall}}{\text{precision}+\text{recall}+\epsilon},
\end{align}
with $\epsilon$ a small constant for numerical stability.

\paragraph{Budget and efficiency.} We log final \texttt{bytes\_used} and the configured \texttt{budget\_bytes}. We report utilization $\texttt{bytes\_used}/\texttt{budget\_bytes}$ and write density $|W|/T$.

\paragraph{Utility-per-KB.} Utility-per-KB normalizes the retained utility by memory consumption (kilobytes), emphasizing low-budget performance.

\paragraph{Drift coverage and staleness.} Drift coverage measures the fraction of labeled drift events that are retained. Average staleness measures how old retained items are at the end of the episode.

\paragraph{Regret (budget-constrained oracle).} Let each step have value $u_t$ and weight $w_t$ given by the benchmark byte estimator. In our implementation, $w_t$ is computed as the length of a canonical JSON serialization of $(x_t, m_t)$ plus fixed per-item overheads (32-byte header + 16-byte index entry). Define the oracle utility under budget $B$ as:
\[
U^\star(B) = \max_{S \subseteq \{1..T\}} \sum_{t\in S} u_t \quad \text{s.t.} \quad \sum_{t\in S} w_t \le B.
\]
We report final regret as $\mathrm{regret}_{\text{write-only}} = \max(0, U^\star_{\text{write-only}}(B) - U)$, clamped to be non-negative for interpretability. The oracle $U^\star_{\text{write-only}}(B)$ is computed under a WRITE-only action space (0/1 knapsack over candidate WRITE events under the same byte estimator). When MERGE is enabled, policies can legitimately exceed this WRITE-only baseline via delta storage; the clamp keeps the reported regret comparable across policies. The implementation uses exact knapsack DP when feasible and a documented greedy approximation otherwise.

\section{Experiments}
\label{sec:experiments}
We evaluate each baseline write policy across drift regimes and memory budgets.

\paragraph{Regimes.} We report results for synthetic drift regimes (Section~\ref{sec:taskgen}): default, burst drift, redundancy, and a combined burst+redundancy regime.

\paragraph{Budgets.} We sweep a range of byte budgets to expose low-budget tradeoffs and high-budget saturation. Unless otherwise stated, budgets are the fixed values used by the evaluation script (and can be modified by editing the sweep list).

\paragraph{Episodes and aggregation.} For each (mode, budget, policy) condition, we run a fixed set of episodes and report the mean of each scalar metric across episodes. We emphasize deterministic evaluation: episodes are generated from fixed seeds and can be frozen to disk so future code changes do not silently change the benchmark instances.

\begin{table}[t]
\centering
\begin{tabular}{ll}
\toprule
Parameter & Default value \\
\midrule
$T$ (timesteps per episode) & 200 \\
\texttt{api\_pool} & 8 \\
\texttt{drift\_prob} & 0.08 \\
\texttt{burst\_interval} & 50 \\
\texttt{burst\_len} & 8 \\
\texttt{burst\_drift\_prob} & 0.6 \\
\texttt{redundancy\_prob} & 0.7 \\
Budgets (bytes) & \{1024, 10240, 102400, 1048576\} \\
Episodes per condition & 10 \\
Seeds & 0 (episode $i$ uses seed $0+i$) \\
\bottomrule
\end{tabular}
\caption{Default generator and evaluation parameters used in our experiments.}
\label{tab:default_params}
\end{table}

\paragraph{Evaluation protocol.} Each run processes the episode stream sequentially. The policy may emit a sequence of memory actions per step (WRITE/MERGE/EXPIRE/SKIP). Actions that violate the byte budget are rejected. EXPIRE is constrained to targets older than the current timestep. After processing the full stream, the final memory contents are scored against the episode labels and utilities.

\paragraph{Implementation note.} The repository includes scripts to reproduce all tables and figures and a small unit-test suite that checks metric invariants and budget-related edge cases.

\section{Results}
\label{sec:results}
We report both (i) task-quality outcomes (F1 and regret) and (ii) memory-behavior diagnostics that explain \emph{why} a policy succeeds or fails under a fixed byte budget.

\subsection{Task quality and budget tradeoffs}
\begin{figure}[t]
\centering
\begin{subfigure}{0.49\linewidth}
  \includegraphics[width=\linewidth]{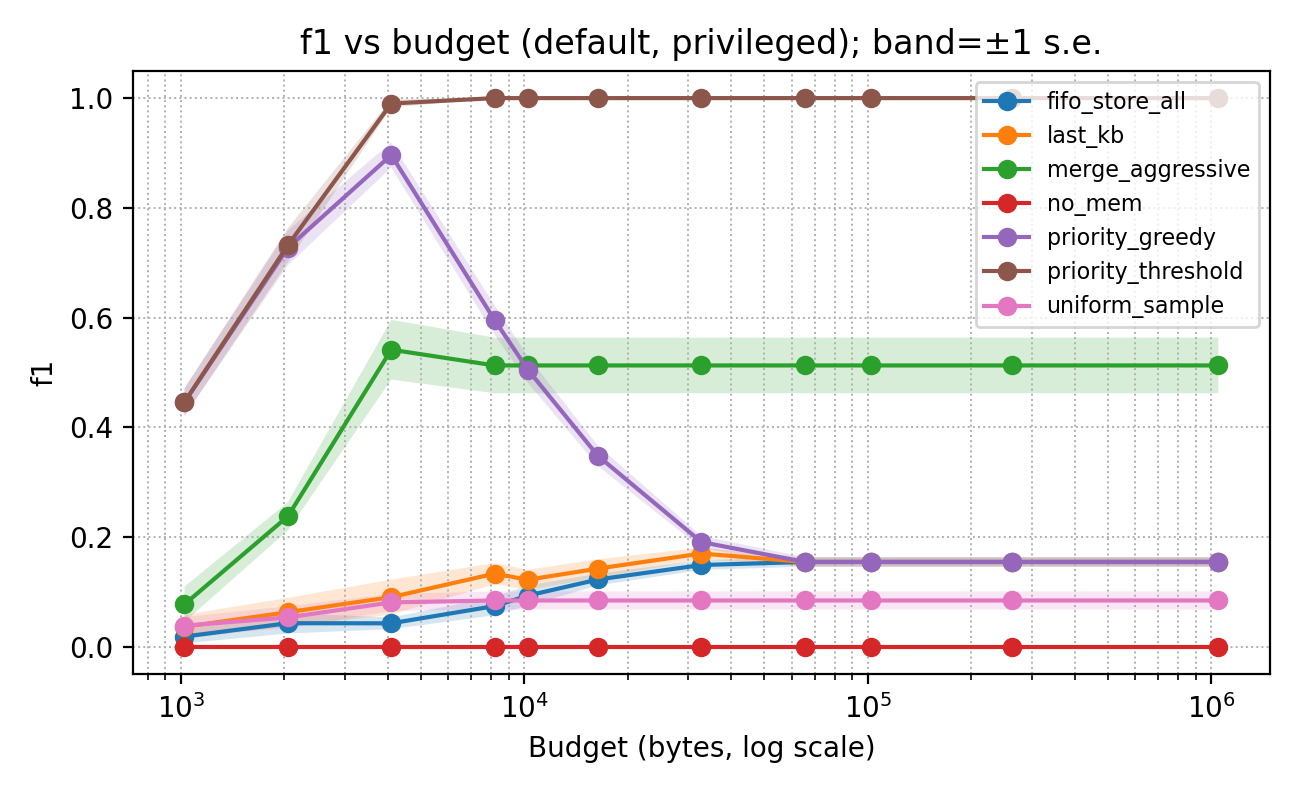}
  \caption{F1 vs. budget (default)}
\end{subfigure}
\begin{subfigure}{0.49\linewidth}
  \includegraphics[width=\linewidth]{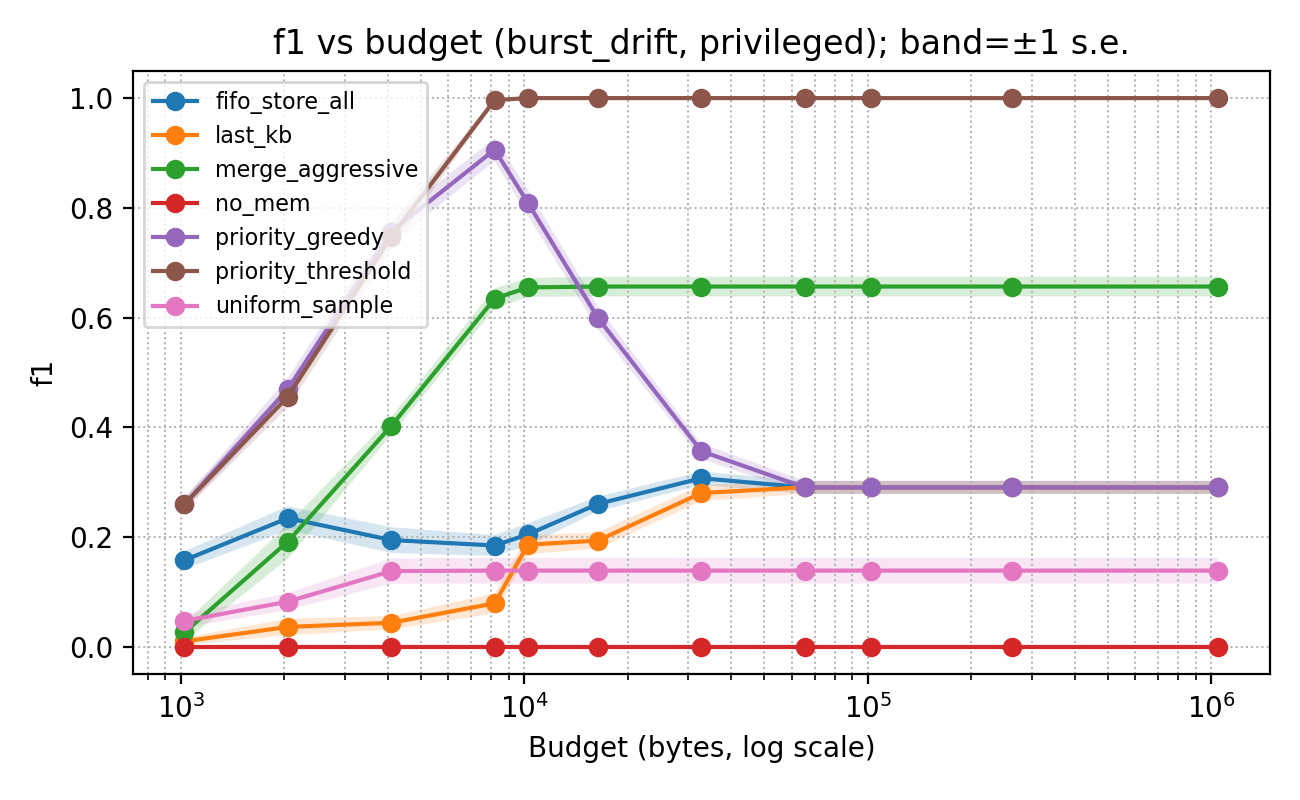}
  \caption{F1 vs. budget (burst drift)}
\end{subfigure}

\vspace{0.5em}
\begin{subfigure}{0.49\linewidth}
  \includegraphics[width=\linewidth]{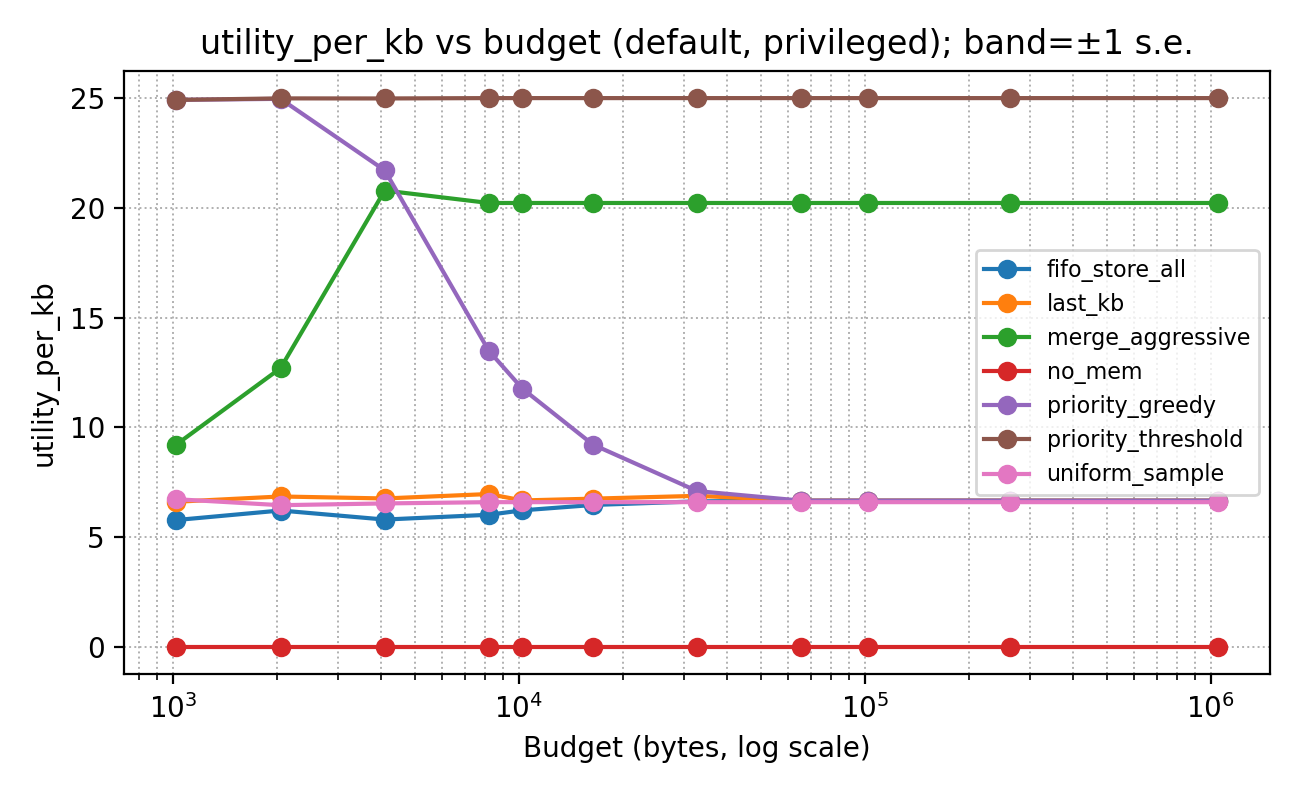}
  \caption{Utility/KB vs. budget (default)}
\end{subfigure}
\begin{subfigure}{0.49\linewidth}
  \includegraphics[width=\linewidth]{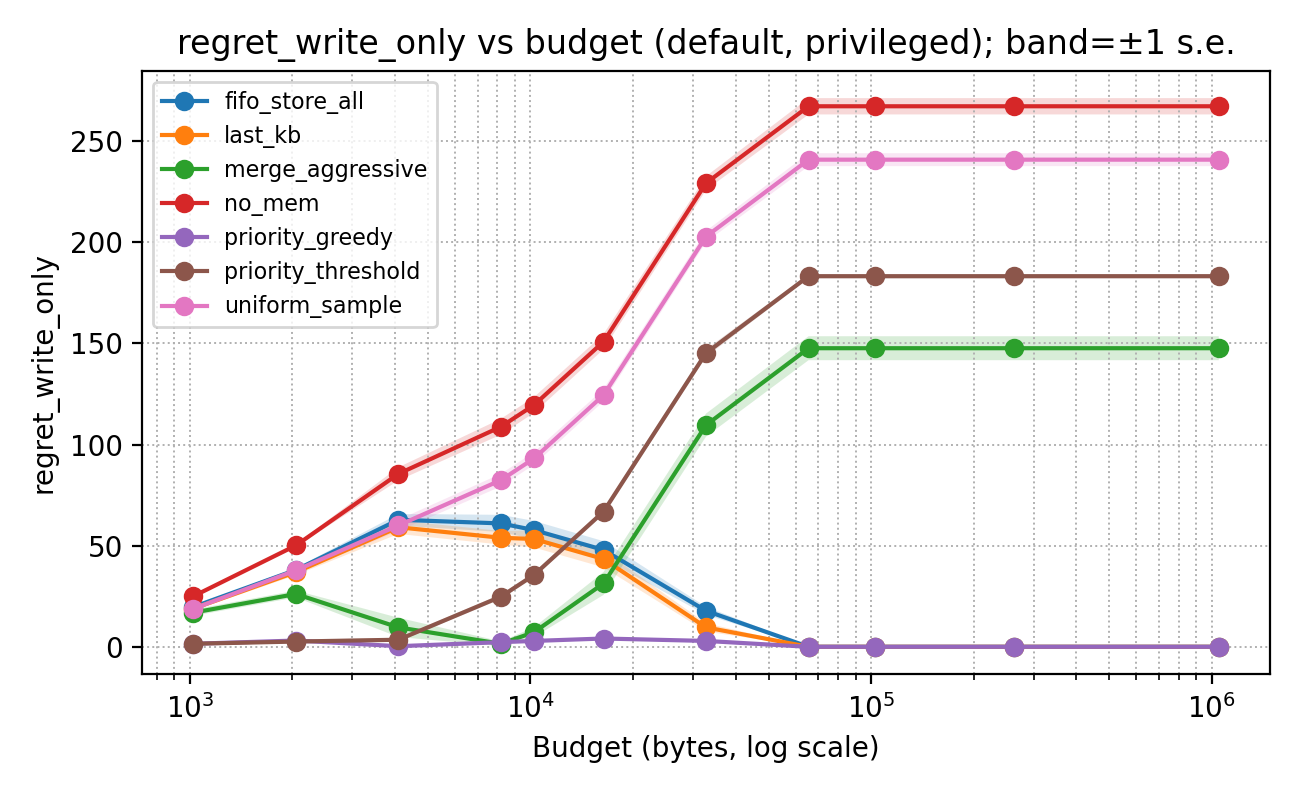}
  \caption{WRITE-only regret vs. budget (default)}
\end{subfigure}
\caption{Budget curves for representative regimes (privileged track). Shaded regions indicate $\pm 1$ standard error over 10 episodes.}
\label{fig:budget-curves}
\end{figure}

We report privileged-track results in the main text for clarity; unprivileged-track curves are available in the repository.
Figure~\ref{fig:budget-curves} shows representative budget--performance curves for the default and burst-drift regimes. Table~\ref{tab:f1_default} reports mean F1 across budgets for the default regime.

\begin{table}[t]
\centering
\begin{tabular}{lcccc}
\toprule
Policy & 1024 & 10240 & 102400 & 1048576 \\
\midrule
\texttt{fifo\_store\_all} & 0.019 & 0.093 & 0.155 & 0.155 \\
\texttt{last\_kb} & 0.036 & 0.122 & 0.155 & 0.155 \\
\texttt{merge\_aggressive} & 0.078 & 0.513 & 0.513 & 0.513 \\
\texttt{no\_mem} & 0.000 & 0.000 & 0.000 & 0.000 \\
\texttt{priority\_greedy} & 0.446 & 0.505 & 0.155 & 0.155 \\
\texttt{priority\_threshold} & 0.446 & 1.000 & 1.000 & 1.000 \\
\texttt{uniform\_sample} & 0.039 & 0.084 & 0.084 & 0.084 \\
\bottomrule
\end{tabular}
\caption{Mean F1 vs. memory budget for the default regime in the \textbf{privileged} track. Budgets are in bytes.}
\label{tab:f1_default}
\end{table}

We additionally report mean F1 for a harder combined regime (burst+redundancy) in Table~\ref{tab:f1_burst_redundancy}.
\begin{table}[t]
\centering
\begin{tabular}{lcccc}
\toprule
Policy & 1024 & 10240 & 102400 & 1048576 \\
\midrule
\texttt{fifo\_store\_all} & 0.126 & 0.176 & 0.285 & 0.285 \\
\texttt{last\_kb} & 0.009 & 0.194 & 0.285 & 0.285 \\
\texttt{merge\_aggressive} & 0.059 & 0.591 & 0.592 & 0.592 \\
\texttt{no\_mem} & 0.000 & 0.000 & 0.000 & 0.000 \\
\texttt{priority\_greedy} & 0.267 & 0.808 & 0.285 & 0.285 \\
\texttt{priority\_threshold} & 0.260 & 1.000 & 1.000 & 1.000 \\
\texttt{uniform\_sample} & 0.034 & 0.108 & 0.108 & 0.108 \\
\bottomrule
\end{tabular}
\caption{Mean F1 vs. memory budget for the combined burst+redundancy regime in the \textbf{privileged} track. Budgets are in bytes.}
\label{tab:f1_burst_redundancy}
\end{table}

\paragraph{Qualitative trends.} Drift coverage/recall saturates once the budget can store essentially all labeled drift events, while F1 may remain low if many non-critical steps are also retained (precision collapse). Some policies degrade at high budgets because increased capacity reduces eviction pressure, allowing more low-value steps into memory; this increases recall but lowers precision and thus F1.

Finally, $\mathrm{regret}_{\text{write-only}}$ is defined against a \emph{budget-dependent} oracle $U^\star_{\text{write-only}}(B)$; as $B$ increases, the oracle can improve, so if a policy's retained utility saturates the optimality gap can grow.

\FloatBarrier
\subsection{Memory behavior diagnostics}
\label{sec:diagnostics}
Task-quality metrics alone can hide distinct failure modes. We therefore report diagnostics derived from the per-step action logs: utilization (fraction of budget used), write density (fraction of timesteps written), expire rate (EXPIRE-to-WRITE ratio), average staleness (age of retained items at episode end), and drift coverage (fraction of labeled drift events retained).

Table~\ref{tab:diagnostics_default_10k} summarizes these diagnostics at a representative low budget (10{,}240 bytes) for the default regime.

\begin{table}[t]
\centering
\begin{tabular}{lccccc}
\toprule
Policy & utilization & write\_density & expire\_rate & avg\_staleness & drift\_coverage \\
\midrule
fifo\_store\_all     & 0.994 & 0.247 & 0.000 & 174.8 & 0.188 \\
last\_kb             & 0.993 & 0.249 & 0.751 &  25.1 & 0.246 \\
priority\_threshold  & 0.337 & 0.084 & 0.000 &  94.3 & 1.000 \\
merge\_aggressive    & 0.589 & 0.262 & 0.000 & 104.8 & 1.000 \\
\bottomrule
\end{tabular}
\caption{Memory diagnostics at 10{,}240 bytes in the default regime (means over episodes; subset of representative policies). High utilization can coincide with low drift coverage when writes are spent on non-critical steps.}
\label{tab:diagnostics_default_10k}
\end{table}

Two patterns recur across regimes. First, \emph{high utilization is not sufficient}: the fill-then-stop baseline (\texttt{fifo\_store\_all}) uses nearly the full budget while achieving only modest drift coverage because it spends many writes on low-value steps. Second, eviction-heavy policies can show very low staleness (e.g., last-$k$) but still miss drift events when bursts occur or when critical steps are not aligned with the eviction window.

\FloatBarrier
\begin{samepage}
\Needspace{8\baselineskip}
\paragraph{MERGE behavior.} The merge-aggressive baseline can improve drift coverage at fixed budgets by compacting multiple updates into fewer memory items, but it is not consistently dominant and can induce brittle behavior when the merge heuristic misidentifies ``same API'' updates. Accordingly, we treat MERGE as optional rather than a central conclusion from the current baseline set.
\end{samepage}

\subsection{Case studies: characteristic failure modes}
\paragraph{Redundancy: wasting bytes on duplicates.} In the redundancy regime, many observations are repeated or near-duplicates. Write-all style policies tend to spend budget on repeated content, driving utilization high without improving drift coverage. By contrast, policies that filter on priority (or that effectively compress repeated updates) maintain higher drift coverage at the same budget because they reserve writes for informative steps.

\paragraph{Burst drift: eviction thrashing and missed clusters.} In the burst-drift regime, critical changes arrive in short clusters. Window-style eviction (last-$k$) reacts by rapidly expiring recent context to accommodate new writes, yielding a high expire rate and low average staleness. However, this ``thrashing'' can still miss drift events when the burst exceeds the effective window size. Priority-aware policies are more robust in this regime because they prioritize retaining high-priority steps during the burst, improving drift coverage and reducing regret under the same byte constraint.

\section{Limitations}
\label{sec:limitations}
This benchmark currently focuses on controlled synthetic drift streams where relevance is labeled by the generator. This design makes evaluation deterministic but does not capture all complexities of real production drift (e.g., ambiguous user goals, noisy logs, and multi-document dependencies).

Our baselines are intentionally simple and do not represent state-of-the-art agentic systems. The goal is to provide a clean substrate for future work on learned write policies, richer regimes (delayed relevance, contradictions), and integration with downstream question answering models.

\section{Related work}
\label{sec:related}
WritePolicyBench sits at the intersection of learning under drift, bounded-memory selection/caching, and memory-augmented systems.

Concept drift and non-stationarity are widely studied in streaming and online learning \cite{gama2014survey}. Cache replacement and eviction strategies provide classic baselines for bounded storage problems \cite{podlipnig2003survey}.

Our ``write under a byte budget'' framing also relates to online/streaming selection problems, including knapsack-style objectives and streaming summarization/subset selection \cite{Kellerer2004Knapsack,IbarraKim1975Knapsack,badanidiyuru2014streaming}.

Memory-augmented architectures motivate explicit external memory as a design axis \cite{graves2016dnc,graves2014ntm}.

Finally, recent LLM-centric systems emphasize external memory beyond raw context windows, typically combining retrieval with memory construction and updates \cite{lewis2020rag,park2023generativeagents,Packer2023MemGPT,Zhong2023MemoryBank}. WritePolicyBench isolates and measures the write/update decision itself under strict byte budgets.

\section{Reproducibility}
\label{sec:repro}
We release code to regenerate all reported numbers and plots.

\paragraph{Repository reference.} The implementation and scripts are available at \href{https://github.com/edgardcham/WritePolicyBench}{\nolinkurl{github.com/edgardcham/WritePolicyBench}} (release tag: \texttt{v1.0}, commit: \texttt{9cfc1f4}).

\paragraph{Determinism.} The benchmark is deterministic given a fixed seed and a frozen set of episodes. We recommend freezing the episode set once (to avoid changes in the generator affecting historical comparisons), then running the evaluator over that frozen set.

\paragraph{Reproducing results.} To reproduce the paper figures and tables, run the provided scripts to: (i) freeze the episode set, (ii) evaluate all (regime, budget, policy) conditions to produce per-episode metric logs, and (iii) aggregate these logs into the plots and LaTeX tables included in this paper.

\paragraph{Unit tests.} The test suite checks basic invariants (e.g., no-memory policies yield zero write density and zero coverage; write-all policies saturate at sufficiently large budgets; staleness and utilization behave as expected). These tests are intended to catch metric regressions and budget-accounting mistakes.

\end{document}